\documentclass[prb,aps,twocolumn,epsfig,showpacs]{revtex4}
\usepackage{graphicx,epsf}
\usepackage{wasysym}
\usepackage{bm}
\usepackage{pstricks}
\usepackage{psfrag}

\newcommand{\beq}{\begin{equation}}
\newcommand{\eeq}{\end{equation}}
\newcommand{\beqn}{\begin{eqnarray}}
\newcommand{\eeqn}{\end{eqnarray}}

\newcommand{\nn}{\nonumber}
\newcommand{\bq}{{\bf q}}

\newcommand{\bK}{{\bf K}}
\newcommand{\bA}{{\bf A}}

\newcommand{\bR}{{\bf R}}
\newcommand{\bp}{{\bf p}}
\newcommand{\br}{{\bf r}}

\newcommand{\nubar}{\bar{\nu}}
\newcommand{\rhobar}{\bar{\rho}}

\newcommand{\etab}{\mbox{\boldmath $\eta $}}

\newcommand{\twd}{two-dimensional}

\def\subsub#1{\noindent{\bf #1:}}

\def\tit#1#2#3#4#5{{#1}{\bf #2}, #3 (#4)}

\def\npb{Nucl.\ Phys.\ B\ }

\def\epjb{Eur.\ Phys.\ J.\ B\ }

\def\prl{Phys.\ Rev.\ Lett.\ }
\def\pr{Phys.\ Rev.\ }
\def\prb{Phys.\ Rev.\ B\ }

\def\natu{Nature\ }

\def\bei{\begin{itemize}}
\def\eei{\end{itemize}}

\begin{document}


\title{Electron interactions in graphene in a strong magnetic field}

\author{M. O. Goerbig,$^1$ R. Moessner,$^2$ and B. Dou\c cot$^3$}

\affiliation{$^1$Laboratoire de Physique des Solides, CNRS UMR 8502,
Universit\'e Paris Sud, Orsay}

\affiliation{$^2$Laboratoire de Physique Th\'eorique de l'Ecole Normale
Sup\'erieure, CNRS UMR 8549, Paris}

\affiliation{$^3$Laboratoire de Physique Th\'eorique et Hautes Energies,
CNRS UMR 7589, Universit\'e Paris 6 et 7, Paris}

\date{\today}

\begin{abstract}
  Graphene in the quantum Hall regime exhibits a multi-component structure due
  to the electronic spin and chirality degrees of freedom. While the applied
  field breaks the spin symmetry explicitly, we show that the fate of
  the chirality SU(2) symmetry is more involved: the leading symmetry-breaking
  terms differ in origin when the Hamiltonian is projected onto the central
  ($n=0$) rather than any other Landau levels. Our description at the
  lattice level leads to a Harper equation; in its continuum limit, the ratio
  of lattice constant $a$ and magnetic length $l_B$ assumes the role of a
  small control parameter in different guises. The leading symmetry-breaking
  terms are lattice effects, 
  algebraically small in $a/l_B$.  We analyze the Haldane
  pseudopotentials for graphene, and evaluate the easy-plane anisotropy
  of the graphene ferromagnet.
\end{abstract}


\maketitle

\subsub{Introduction} The recent discovery \cite{novoselov1,zhang1} of
an integer quantum Hall (QH) effect in a \twd~(2D) sheet of graphite, known as
graphene, has triggered an avalanche of activity, 
including on the theory side  studies of the transport properties of
relativistic Dirac particles,\cite{gusynin,katsnelson,peres}
the analysis of edge states \cite{CN,brey,abanin} and shot noise 
\cite{bjoern} as well as  Berry phases in bilayers.\cite{mccann}

In a simple model, electrons in graphene can be treated as hopping on
a honeycomb lattice.\cite{wallace,slon} 
Perhaps the most salient feature of this problem
is the existence in the band structure of a pair of Dirac points with
a linear (`relativistic') energy-momentum relationship. These points
are located at the two inequivalent corners of the Brillouin zone
(labelled by $K$ and $K'=-K$), endowing graphene with a
multi-component structure analogous to the well-studied examples of
bilayer systems, the multi-valley structure of Silicon and of course
the simple spin degree of freedom. These degrees of freedom can be
thought of as SU(2) (or higher symmetry) pseudospins. The resulting
Hamiltonian typically contains symmetric terms as well as ones which
lower the symmetry, such as Zeeman (spin) or capacitance (bilayer)
\cite{moon} energies.

Here, we argue that graphene may be viewed as a further type of
multi-component system.  Its internal degree of freedom can be thought of as a
chirality:\cite{haldane} the wavevectors $K$ and $K'$ encode the
(anti)clockwise variation of the phase of the electronic wavefunction on the
three sites neighboring any given site on one sublattice.

Moreover, we show that for $n\neq 0$
the chiral SU(2) symmetry is reduced to U(1) in graphene, due to
backscattering
terms with momentum transfer $2K\sim K$ which  
provide a coupling of the chirality
to the orbital part of the wavefunction. On the contrary,
in 
$n=0$, the broken symmetry may be due to 
electrostatic (``Hartree'') effects.  Although different in origin,
both effects are of order $O(a/l_B)$.
The distance between neighboring carbon atoms $a\simeq 0.14 {\rm nm}$
provides an additional lengthscale besides the magnetic length
$l_B=\sqrt{\hbar/eB}
=26{\rm nm}/\sqrt{B[{\rm T}]}$, so that $a\ll l_B$. 
It is somewhat analogous to
the layer separation $d$ for bilayers. In graphene for $n\neq0$, however, only
the exchange part of the symmetry-breaking interaction is non-zero, whereas in
bilayers the direct term encodes the capacitance energy, which can be
important already for typical values of $d/l_B\sim 1$. 

In the following, we flesh out this picture with a microscopic
calculation starting at the lattice level, 
in which we derive and discuss the effective model for interacting
electrons restricted to a single relativistic LL and compare
it to the non-relativistic case of electrons in conventional
semiconductor heterostructures; the difference between the two is
most significant 
for $n=1$. The backscattering terms are discussed in the case
of the QH ferromagnet at the filling factors $\nubar=1$ of
the partially filled LL (for an arbitrary LL, we have
$\nu=4n+\nubar$). 

\subsub{The model} 
The electron field in graphene may be written as a
two-spinor whose components,
$\psi_{\sigma}(\br)=\exp(i\sigma\bK\cdot\br)\chi_{\sigma}(\br)$, are a
product of a slowly varying part $\chi_{\sigma}(\br)$ and a rapidly
oscillating plane wave with 
$\sigma\bK=
\sigma(4\pi/3\sqrt{3}a){\bf e}_x$ for the Brillouin
zone corners $K$ and $K'$ (chiralities $\sigma=\pm1$,
respectively).
The components of each two-spinor field $\chi_{\sigma}(\br)$
correspond to the two triangular sublattices 
(labelled by $\alpha=\pm1$)
of the bipartite honeycomb lattice.
In a magnetic field with the Landau gauge $\bA=(0,Bx)$, 
$q_y$ is a good quantum number, and we may expand $\chi_{\alpha}(\br)=
\exp(iq_yy)g_{\alpha}(y)$. The electron dynamics (in a tight-binding model with
nearest-neighbor hopping $t=1$) is governed by the Harper equation

\beqn\label{equ0001}
\nn
Eg_{\alpha}^{\sigma}(x)&=&-2\cos\left\{\sigma\frac{2\pi}{3}+
\frac{\sqrt{3}}{2}\left[q_y+
\frac{\left(x+\alpha/4\right)}{l_B^2}\right]\right\}\\
&&\times g_{-\alpha}^{\sigma}\left(x+\frac{\alpha}{2}\right)-g_{-\alpha}^{\sigma}(x-\alpha),
\eeqn
where the distances are measured in units of $a$. In order to
derive a continuum limit in the presence of an unbounded vector potential
$\bA$, one expands  the cosine in Eq.~(\ref{equ0001}) in the 
vicinity of $x_\mu$ defined as
$[q_y+\alpha x_\mu(q_y)/l_B^2]\sqrt{3}/2=2\pi \mu$, 
where $\mu$ is an integer which effectively acts as an additional 
quantum number besides the quasimomentum $q_y$ in the first Brillouin zone.

The continuum limit of Eq. (\ref{equ0001}) thus reads
$$Eg_{\alpha}^{\sigma}(x)=\frac{3}{2}\left(\alpha l_B\partial_x+\sigma\frac{x}{l_B}\right)
g_{-\alpha}^{\sigma}(x),$$
where $x$ is now a small deviation from $x_{\mu}$. This result coincides with
the ones obtained by introducing the minimal coupling $\bp\rightarrow
\bp+e\bA$ {\sl after} deriving the $B=0$ continuum theory. 
\cite{haldane,LLquant} 
Note that the typical extension of the wavefunctions along the $x$-axis
is $R_L\propto\sqrt{n}l_B$ in the $n$-th LL, and
the periodicity of $x_{\mu}$ is $\sim l_B^2/a$. 
The overlap between wavefunctions with differing $\mu$ is therefore
{\em exponentially suppressed 
provided $\sqrt{n}\ll l_B/a$}.
Finally,
\beqn\label{equ002}
\chi_+(\br)=
\frac{1}{\sqrt{2}}
\sum_{n,m}
\left(\begin{array}{c}
i\sqrt{1+\delta_{n,0}}
\langle 
\br|\,|n|,m\rangle \\ {\rm sgn}(n)
\langle \br|\,|n|-1,m\rangle \end{array}\right)c_{n,m;+}\ ,
\eeqn
where the index $+$ 
represents the $K$ point. In the expression for $\chi_-$
(at $K'$), the components of the spinor are reversed.
Here ${\rm sgn}(n)=\{1,0,-1\}$ for $n\{>,=,<\}0$, respectively.
The quantum number $n$ is the index of the relativistic LL, 
and $m$ is associated to the guiding center operator, 
which commutes with the one-particle Hamiltonian. The $|n,m\rangle$ are
the usual (non-relativistic) one-particle states for a charged particle in
a perpendicular magnetic field. The $c_{n,m;\sigma}$ are fermionic
destruction operators.

Projection onto a LL ($n\neq0$) of the  sublattice densities
$\rho_{\alpha}(\br)=\sum_{\sigma,\sigma'}
\psi_{\sigma,\alpha}^{\dagger}(\br)\psi_{\sigma',\alpha}(\br)
$
gives
\beq\label{equ003}
\rho^n(\bq)=\rho_1^{n}(\bq)+\rho_2^{n}(\bq)
=\sum_{\sigma,\sigma'}F_n^{\sigma\sigma'}
(\bq)\rhobar^{\sigma\sigma'}(\bq),
\eeq
where the {\sl projected density operators} read
$\rhobar^{\sigma\sigma'}(\bq)=\sum_{m,m'}\langle m|\exp\{
-i[\bq+(\sigma-\sigma')\bK]\cdot\bR\}|m'\rangle 
c_{n,m,\sigma}^{\dagger}c_{n,m',\sigma'}$.
The operator
$\bR$ represents the usual guiding center position, and $\etab$ is the 
cyclotron coordinate, $\br=\bR+\etab$. The
chirality-dependent form factors $F_n^{\sigma\sigma'}(\bq)$ read,
in terms of
associated Laguerre polynomials $L_n^{\alpha}(x)$: 
\beqn\label{equ006}
\nn
F_n^{++}(\bq)&=&\frac{1}{2}\left[
L_{|n|}\left(\frac{|\bq|^2}{2}\right)
+L_{|n|-1}\left(\frac{|\bq|^2}{2}\right)\right]e^{-|\bq|^2/4},\\
\nn
F_n^{+-}(\bq)&=& \left(\frac{-(q+q^*-K-K^*)}{2\sqrt{2|n|}}\right)\\
\nn
&&\times
L_{|n|-1}^1\left(\frac{|\bq-\bK|^2}{2}\right)e^{-|\bq-\bK|^2/4},
\eeqn
where $q=q_x+iq_y$ and $K=K_x+iK_y$ are written in complex notation, and 
the wave vectors are given in units of $1/l_B$.
$F_n^{++}(\bq)=F_n^{--}(\bq)$, and $F_n^{-+}(\bq)$ is obtained by
replacing $\bq\to-\bq$ in $F_n^{+-}(\bq)$.

The Hamiltonian of interacting electrons in graphene, projected onto a single
relativistic LL thus reads
\beq\label{equ007}
H=\frac{1}{2}\sum_{\sigma_1,...,\sigma_4}\sum_{\bq}v_n^{\sigma_1,...,\sigma_4}
(\bq)\rhobar^{\sigma_1\sigma_3}(-\bq)\rhobar^{\sigma_2\sigma_4}(\bq),
\eeq
where the sum over the wave vectors is restricted to the first Brillouin zone.
Indeed, the potential consists of a sum over reciprocal lattice vectors,
as the local densities [Eq. (\ref{equ003})], valid for
$l_B\gg a$, are restricted to the hexagonal lattice. 
The chirality-dependent effective interaction
\beq\label{equ008}
v_n^{\sigma_1,...,\sigma_4}(\bq)=\frac{2\pi e^2}{\epsilon|\bq|}
F_n^{\sigma_1\sigma_3}(-\bq)F_n^{\sigma_2\sigma_4}(\bq)
\eeq
is not SU(2) symmetric;
however, the symmetry-breaking terms are suppressed parametrically in 
$a/l_B$. To see this, consider 
the different form-factor combinations in the
effective interaction potential (\ref{equ006}).\\
-- Terms of the form $F_n^{\sigma,\sigma}(\mp \bq)F_n^{\sigma',-\sigma'}
(\pm\bq)$ and ``umklapp scattering'' terms 
[$F_n^{\sigma,-\sigma}(-\bq)F_n^{\sigma,-\sigma}(\bq)$] are 
{\sl exponentially small} in $a/l_B$.\\
-- ``Backscattering''
[$F_n^{\sigma,-\sigma}(-\bq)F_n^{-\sigma,\sigma}(\bq)$]: 
one obtains
$v_n^{+--+}(\bq)\sim \exp(-|\bq|^2/2)/|\bq'\pm\bK|
\sim \exp(-|\bq|^2/2)/|\bK|,$
which is only {\sl algebraically small},
$v_n^{+--+}/(e^2/\epsilon l_B)\sim a/l_B$,
and thus constitutes the leading perturbation to the  
remaining [SU(2) invariant] terms.

These leading-order terms in the effective interaction 
yield the SU(2) [or SU(4), if the physical spin is also taken into account]
symmetric Hamiltonian (for $n\neq 0$)
\beq\label{equ009}
H_{eff}^n=\frac{1}{2}\sum_{\sigma,\sigma'}
\sum_{\bq} \frac{2\pi e^2}{\epsilon |\bq|}\left[
\mathcal{F}_n(q)\right]^2\rhobar_{\sigma}(-\bq)\rhobar_{\sigma'}(\bq),
\eeq
with the {\sl graphene form factor}
\beq\label{equ010}
\mathcal{F}_n(q)=\frac{1}{2}\left[L_{|n|}\left(\frac{q^2}{2}\right)+
L_{|n|-1}\left(\frac{q^2}{2}\right)\right]e^{-q^2/4}
\eeq
and $\rhobar_{\sigma}(\bq)\equiv \rhobar^{\sigma\sigma}(\bq)$.
The graphene form factor 
(\ref{equ010}) has already been written down by Nomura and MacDonald
in their 
study of the QH ferromagnetism at $\nubar=1$.\cite{macdonald}
The leading-order symmetry-breaking
correction due to backscattering is (with 
$v_n^{\sigma,-\sigma}(\bq)\equiv v_n^{\sigma,-\sigma,-\sigma,\sigma}(\bq)$)
\beq\label{equ012}
H_{bs}=\frac{1}{2}\sum_{\sigma}\sum_{\bq}v_n^{\sigma,-\sigma}(\bq)
\rhobar^{\sigma,-\sigma}(-\bq)\rhobar^{-\sigma,\sigma}(\bq).
\eeq

The central level ($n=0$) behaves remarkably differently. In this
case, the electron chirality $\sigma$ is equivalent to the sublattice index
$\alpha$ (Eq.~\ref{equ002}), and therefore
$$\rho_{\alpha}^{n=0}(\bq)=e^{-q^2/4}
\sum_{m,m'}\langle m|e^{-i\bq\cdot\bR}|m'\rangle
c_{n,m;\alpha}^{\dagger}c_{n,m';\alpha}$$
with the same form factor 
$\mathcal{F}_{n=0}(q)=e^{-q^2/4}$
as for non-relativistic electrons in the lowest LL. From an electrostatic point
of view, it may be energetically favorable to distribute the electronic 
density with equal weight on both sublattices. For $n\neq
0$, this follows directly from Eq. (\ref{equ002}), but in $n=0$, an
equal-weight superposition of $\sigma=\pm1$ is required to distribute the 
charges homogeneously on both sublattices.
Such an electrostatic effect, compared to the SU(2) invariant
terms, is of the same order $O(a/l_B)$ as the backscattering term in $n\neq 0$.

\begin{figure}
\centering
\includegraphics[width=6.5cm,angle=0]{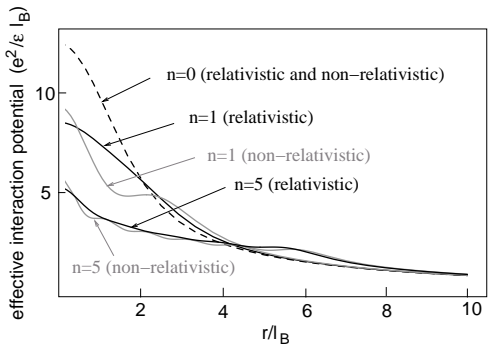}
\caption{\footnotesize{Effective interaction potentials in real space: 
comparison between the relativistic (black lines) and non-relativistic
(gray lines) LLs $n=0,1$, and $5$. The dashed line represents the interaction
potential in $n=0$, which is the same for relativistic and 
non-relativistic electrons.
}}
\label{fig01}
\end{figure}

\subsub{Effective interaction} 
By absorbing the form factor into the interaction, we 
define
\beq\label{equ015}
v_n(q)=\frac{2\pi e^2}{\epsilon q}\left[\mathcal{F}_n(q)\right]^2.
\eeq
Fig.\,\ref{fig01}
shows the effective interaction
potentials (\ref{equ015}) transformed to real space, 
for $n=0,1$, and $5$. At large distances, the usual $1/r$ Coulomb potential is 
obtained. Interestingly, the shape of the interaction potential for the
relativistic $n=1$ LL in graphene is more similar to the non-relativistic
$n=0$ level than to the corresponding one $n=1$, as may also be seen in 
a pseudopotential expansion,\cite{haldane2}
$V_{\ell}^n=(2\pi)^{-1}\sum_{\bq}v_n(q)L_{\ell}(q^2)\exp(-q^2/2).$
Indeed the ratios $V_{2m+1}/V_{2m+3}$ for the odd integer pseudopotentials, 
which are relevant for the case of polarized electrons, decrease monotonically
both in $n=1$ and $n=0$ for relativistic LLs. These ratios, and the differences
$V_{2m-1}-V_{2m+1}$ are bigger in the former case, so that -- among the
polarized states -- fractional QH states will therefore be more stable in
$n=1$ than in $n=0$ (at constant magnetic field). By contrast, candidate
chirality unpolarized states (such as at $\nu=2/3$) fare better, for two
reasons: firstly, the fact that the relativistic effective potential is more
short-ranged in $n=0$ than in $n=1$ leads to $V_0$ (and $V_0/V_1$) being
smaller for $n=1$. Secondly, the pair of internal SU(2) degrees of freedom
allow for a smaller unpolarized 'composite Fermi sphere'.

Numerical results \cite{RH} show a first-order phase transition at
$V_1/V_3\simeq 1.3$ between the Pfaffian state \cite{MR} at $\nubar=1/2$ and a
charge-density wave,\cite{CDW} and a crossover to a composite-fermion Fermi
sea when $V_1/V_3$ is further increased. The Pfaffian state is absent in
$n=0$, where $V_1^0/V_3^0=1.6$, probably due to an inaccessibly small gap.
\cite{RH} In the relativistic $n=1$ LL, one finds an even larger ratio
$V_1^1/V_3^1=1.67$ so that a Pfaffian state is also unlikely to be observed
there. Even though the ratio $V_1^2/V_3^2=1.16$ in the relativistic $n=2$ LL
is larger than in the corresponding non-relativistic level
($V_1^2/V_3^2=1.14$), it is well below the critical ratio, and one would
thus expect a stripe phase at $\nubar=1/2$.

It is straightforward to check that the 
difference to the non-relativistic case vanishes, in the large-$n$
limit (see Fig.~\ref{fig01} for $n=5$), i.e.\ far from the Dirac points.
Replacing $L_n(q^2/2)\exp(-q^2/4)\simeq J_0(q\sqrt{2n+1}),$ the envelope 
of 
$\mathcal{F}^2_n(q)\simeq \left[J_0(q\sqrt{2n-1})+J_0(q\sqrt{2n+1})
\right]^2/4\simeq J^2_0(q\sqrt{2n})$ agrees to leading
order in $n$ with the non-relativistic case.  

\subsub{QH ferromagnet at $\nubar=1$} In recent transport measurements on a
single graphene sheet additional integer QH plateaux beyond those 
corresponding to $\nu=4n$ have been observed.\cite{zhang2} 
These appear as the 
first signature of electron-electron interactions, and the analogy
with the non-relativistic case in semiconductor heterostructures hints at 
a chirality QH ferromagnet. The stability of such a state, 
in the presence of
impurities, has been investigated by Nomura and MacDonald.\cite{macdonald}
We now analyze the impact of the backscattering term (\ref{equ012}) on
such a ferromagnet for $n\neq 0$, within the Hartree-Fock (HF) approximation. 
Following Ref.~\onlinecite{moon}, we consider the HF trial state
$
|\Psi\rangle=\prod_{m}(u_mc_{m,+}^{\dagger}+v_mc_{m,-}^{\dagger})
|0\rangle,
$
where we may parametrize $u_m=\cos(\theta_m/2)e^{-i\phi_m/2}$ and
$v_m=\sin(\theta_m/2)e^{i\phi_m/2}$, in terms of the real angle fields
$\theta_m$  and $\phi_m$, which can be thought of as polar coordinates 
of a vector field ${\bf n}(m)$.
In the case of a SU(2)-symmetric 
repulsive interaction, it has been shown that the trial state $|\Psi\rangle$
minimizes the energy for constant $\theta_m$ and $\phi_m$, thus yielding a
simple ferromagnet.\cite{moon}
The backscattering term, averaged over this state,
is, apart from an unimportant constant $C$,
\beqn\label{equ017}
\nn
\langle H_{bs}\rangle&=&\frac{1}{4}
\sum_{m,m'}\left\{V^{bs}_{m,m',m,m'}
\left[n_x(m)n_x(m')+x\rightarrow y\right]\right.\\
&&\left.+V^{bs}_{m,m',m',m}n_z(m)n_z(m')\right\}+C,
\eeqn
\beqn\label{equ018}
\nn
V_{m_1,...,m_4}^{bs}&\simeq&\frac{\pi e^2}{\epsilon |\bK|}\sum_{\bq}
\frac{|\bq|^2}{2|n|}
\left[L_{|n|-1}^1\left(\frac{|\bq|^2}{2}\right)e^{-|\bq|^2/4}\right]^2\\
&&\times\langle m_1|e^{i\bq\cdot\bR}|m_3\rangle
\langle m_2|e^{-i\bq\cdot\bR}|m_4\rangle.
\eeqn
The factor of $|q|^2$ in this sum is due to the fact that the wavefunctions on
the same sublattice, but for different chiralities, are orthogonal. 
A gradient expansion \cite{moon} yields to lowest order an easy-plane
anisotropy $\Delta_z$:
\beq\label{equ019}
\langle H_{bs}\rangle^{(0)}=\sum_{m}\Delta_z[n_z(m)]^2,~~~
\Delta_z=
\frac{1}{16\pi^2}
\frac{e^2}{\epsilon |\bK|}.\label{equ020}
\eeq
This
is reminiscent of the bilayer case, where a finite layer separation also
induces easy-plane ferromagnetism. The key differences are: (i) the parameter
$a/l_B\sim 10^{-2}$, 
which mimicks the ``layer separation'', is tiny for currently 
experimentally accessible magnetic fields.
This implies a Curie temperature $\Theta\sim e^2/k_B\epsilon
l_B$, whereas the crossover to easy-plane behavior does not become visible 
until a 
logarithmically (in $a/l_B$) small energy. As chirality
ferromagnetism involves neither electric nor magnetic dipole ordering,
inter-plane coupling in a multi-layer system will be suppressed.
This opens the perspective of probing the 2D behavior for instance
in specific-heat measurements. 
(ii) Contrary to the bilayer case
and the relativistic $n=0$ LL,
the gap is not due to a charging energy when only one layer is filled -- 
there is no contribution to Eq.~(\ref{equ020}) from the direct interaction 
because $v^{\pm}_n(q=0)=0$ [Eq.~(\ref{equ008})].
(iii) $\Delta_z$ is a lattice effect -- it vanishes linearly in $a$ 
as the lattice constant tends to zero at fixed $l_B$. It
does not depend on $n$, whereas the SU(2) symmetric terms
scale as $e^2/\epsilon \sqrt{n}$ in the large-$n$ limit. 
Note, however, that the continuum limit based on the Dirac equation 
ceases to be valid when 
$R_L\sim \sqrt{n}l_B\sim l_B^2/a$.

\subsub{Comparison with experiment} Zhang {\sl et al.} have observed
additional Hall plateaux corresponding to $\nu=0,\pm1\ (n=0)$ and $\nu=\pm4\ 
(n=1)$.\cite{zhang2} The former pair corresponds to a complete resolution of
the fourfold degeneracy of LLs corresponding to different internal (spin and
chirality) degrees of freedom in $n=0$. An explanation of this has to consider
the size of the disorder
broadening of the LLs, $\Gamma$, compared to their splitting due to
the cost of exciting quasiparticles away from integer
filling.\cite{macdonald} An experimental estimate yields $\Gamma\sim 1.7$meV.
\cite{zhang2}

Using our above results, we find that these
quasiparticles are Skyrmions for $n=0,1$, whose energy cost 
$E_{sk}=4\pi \rho_s$
is obtained within the non-linear sigma model,\cite{moon}
with the help of the stiffness 
$$\rho_s=\frac{1}{32\pi^2}\int_0^{\infty}dq q^3 v_n(q).$$
One obtains for the experimentally relevant parameters
(at $17$T with dielectric constant\cite{diel} $\epsilon\sim 5$)
$E_{sk}=4\pi \rho_s=\frac{7}{64}\sqrt{\pi/2}e^2/\epsilon l_B\sim
1.8$meV ($n=1$) and, for $n=0$, $E_{sk}=\frac{1}{4}\sqrt{\pi/2}e^2/\epsilon
l_B\sim 4$meV, both for SU(2) or SU(4).\cite{arovas} 
In addition, there is a
contribution from anisotropies, mainly the Zeeman effect (about
$E_Z=0.1B$[T]meV); the chirality-symmetry breaking due to lattice effects,
being of order $\sim 0.05$meV, play only a minor role here.

The activation gap at $\nu=\pm 4$ scales linearly with $B$, indicating a
relevant Zeeman effect, and the plateau is visible from $\sim 17$T onwards.
\cite{zhang2}
Given the Skyrmions in $n=0$ are more costly than the sum of $E_{sk}$ and
$E_Z$ in $n=1$, this explains why the chirality Landau levels are resolved at
17T in $n=0$, even without the help of an anisotropy field, whereas they
remain absent at $\nu=\pm3,\pm5$ in fields up to 45T. In fact, for $n=1$,
$E_{sk}$ does not reach 4meV for fields below 80T; also, the plateau at
$\nu=0$ disappears below $11$T, where $E_{sk}\sim3$meV.

To summarize, we have analyzed a microscopic model for interaction effects in
graphene in the QH regime. We find corrections to the SU(2)
chirality-symmetric model to be numerically much smaller than the Zeeman
energy breaking the SU(2) spin-symmetry. In addition, the effective
interaction potential differs from the non-relativistic case most strongly for
small but nonzero $n$, in particular $n=1$, which will therefore a good place 
to look for interaction physics different from the GaAs
heterostructure. Finally, recent experiments suggest the presence of
chirality ferromagnetism and Skyrmions in graphene.

\subsub{Note added} After submission of this manuscript, articles of 
related work appeared, by Alicea and Fisher \cite{AF} on ferromagnetism at the
integer QHE, and by Apalkov and Chakraborty \cite{AC} on exact
diagonalisations in the fractional QH regime using the above-mentioned
pseudopotentials.

\subsub{Acknowledgements} We thank N. Cooper, J.-N. Fuchs, D. Huse, P.
Lederer, and S. Sondhi for fruitful discussions. M.O.G.\ is greatful for a
stimulating interaction in the LPS' ``Journal Club Mesoscopic Physics''.

\end{document}